\documentstyle[psfig,aaspp4,flushrt,12pt]{article}

\newcommand\approxlt{\mbox{$^{<}\hspace{-0.24cm}_{\sim}$}}

\begin{document}
\oddsidemargin 0in
\evensidemargin 0in
\topmargin 0.5in
\title{%
%
Constraining the Power Spectrum Using the Column Density
Distribution: a Status Report
%
} 
%
\author{%
Lam Hui\footnote{
NASA/Fermilab Astrophysics Center, Fermi National Accelerator
Laboratory, Batavia, IL 60510, U.S.A.
}
}
\begin{abstract}
We review the arguments given in [\cite{hgz97}] for how the
slope of the column density distribution of the Ly$\alpha$ forest
should depend on the matter power spectrum. The latest progress,
presented by various groups in this conference and elsewhere, is summarized. 
\end{abstract}
\section{Introduction}
One of the results coming out of the first hydrodynamic simulations of the
low column density Ly$\alpha$ forest (${\rm N_{\rm HI}} \approxlt 10^{14.5} \,
{\rm \, cm^{-2}}$) is that the predicted column density distribution
(we will denote by $f(N_{\rm HI})$)
roughly agrees with observations ([\cite{cmor94}],[\cite{zan95}],
[\cite{hkwm95}] \& [\cite{mcor96}]). Because of the limited number of
cosmological  
models simulated then, it was unclear if the agreement is coincidental or 
a generic prediction of a large class of models.
Subsequent analytic/semi-analytic work using the lognormal approximation 
for the density field ([\cite{bd97}] \& [\cite{gh96}]) predicts that
$f(N_{\rm HI})$'s for a large class of models are very similar. 
It is later shown by [\cite{hgz97}] that 1. the lognormal approximation 
does not 
predict $f(N_{\rm HI})$ accurately when compared with the result 
of a hydrodynamic simulation, although it provides useful physical insights; 2. the Zel'dovich approximation (ZA hereafter), which has been widely tested and used in studies of large scale structure 
[\cite{cms93}], works better. In addition, it is pointed out
that the slope of $f(N_{\rm HI})$ depends on both
the normalization and slope of the power spectrum, and,
using the ZA, it is predicted that
cosmological models with less small scale power would
have a steeper $f(N_{\rm HI})$. 

We review arguments that lead to this conclusion, discuss the possible
pitfalls, and summarize the latest progress.


\section{Less Power Implies Steeper $f(N_{\rm HI})$}

One starts with the following picture of the forest: each Ly$\alpha$ line is 
associated with a local maximum or a peak in the neutral hydrogen density.
The column density for a given peak is:
\begin{equation}
N_{\rm HI} = \int_{\rm peak} n_{\rm HI} dr = \left[n_{\rm HI} 
L\right]_{\rm peak}
\label{nhi}
\end{equation}
where $r$ is the proper distance along the line of sight, $n_{\rm HI}$ is
the proper density of ${\rm HI}$, and the symbol $[\, ]_{\rm peak}$
denotes evaluation at the peak. $L$ is the
width of the peak defined by $L \equiv \sqrt{{-2\pi/[d^2 {\rm ln} \, n_{\rm HI}/d r^2]}}$. The above equation 
can be derived by replacing $n_{\rm HI}$ under the integral
by ${\rm exp}[{\rm ln}\, n_{\rm HI}]$ and performing a Gaussian integration 
after Taylor expanding ${\rm ln}\, n_{\rm HI}$ to second order.

$n_{\rm HI}$ is proportional to the
underlying baryon or gas density $\rho$ raised
to some power, which depends on the reionization history [\cite{hg96}].
$N_{\rm HI}$ is then a simple function of the peak height
in $\rho$ and its second derivative $d^2\rho/dr^2$ which determines the
peak width. Hence, $f(N_{\rm HI})$, which is
the number of absorption lines per unit $N_{\rm HI}$ per unit redshift, 
is a statistic of density peaks: the number density of peaks
with a given combination of peak height and width.

Such a statistic can be computed using the method of [\cite{bbks86}]. 
It turns out that for the low column density 
Ly$\alpha$ forest (${\rm N_{\rm HI}} \approxlt 10^{14.5} \,
{\rm \, cm^{-2}}$), the relevant overdensity $\delta\rho/\bar\rho$ is 
of the order of unity. The ZA is known to work well for such a 
quasilinear density field, and it can be shown generally that the Gaussian nature 
of the ZA displacement field implies the slope of $f(N_{\rm HI})$ 
depends on the normalization and slope of the 
primordial power spectrum [\cite{hgz97}]. We will not discuss the
dependence on the power spectrum slope in detail here,
because its effect is smaller than that of the normalization,
and also because the arguments leading to it are more intricate.
The interested reader is referred to [\cite{hgz97}].

The effect of the power spectrum normalization is simple to understand:
let us compare two models, $A$ and $B$, with $A$ having more power on
the relevant scales (see next sec.) than $B$.
{\bf Model $B$, because it has less power, and so has
a less nonlinear density field, would have proportionally fewer high 
peaks compared to low peaks.} Now, $N_{\rm HI}$ depends on both the
peak height and width (eq. \ref{nhi}), but it turns out that
the peak width is correlated with peak height
in such a way that a larger peak height also means higher $N_{\rm HI}$. 
Therefore, $B$, having fewer high peaks compared to low peaks, also
has fewer high $N_{\rm HI}$ lines compared to low ones, hence
a steeper $f(N_{\rm HI})$. 
We illustrate this in Fig. 1 where the CDM model plays 
the role of $A$ and the CHDM model that of $B$. Also shown in the plot is 
a comparison of the ZA $+$ peak-counting prediction with 
the result of hydrodynamic-simulation $+$ Voigt-profile-fitting for  
the CDM model. The agreement is encouraging. 

\section{Discussion}

Let us examine a few caveats to the above arguments, and
discuss some recent work which has bearing on these issues.
First, the peak-counting is done in real space, whereas the absorption
lines are observed in velocity space. A plausible support for our procedure 
comes from a test in [\cite{hgz97}]
where $f(N_{\rm HI})$'s are computed for two sets of spectra, one 
generated with non-zero peculiar velocities, and the other 
with vanishing ones. They agree with each other well. Moreover, 
one can see from eq. (\ref{nhi}) that for an isolated peak, the velocity structure  
plays no role in determining $N_{\rm HI}$. It is interesting to note that there is a recent paper in which peak-counting 
is done directly in velocity space [\cite{hr97}]. 

Second, when using the ZA, smoothing has to be applied to the initial 
displacement field to minimize the amount of orbit-crossing by
the time of interest. There is a well-tested prescription for
the optimal dark matter smoothing scale, which we call the 
orbit-crossing scale [\cite{cms93}].  
Another relevant smoothing scale is roughly the Jeans length, which 
specifies the scale
on which the baryon density is smoothed with respect to the
dark matter density [\cite{gh97}]. It turns out that both are of the
order of $k^{-1} \sim 0.1 h^{-1} {\rm Mpc}$, which makes the Ly$\alpha$ 
forest interesting, because it constrains the power spectrum on these 
scales which are smaller than those probed by galaxy surveys. 
The procedure adopted in [\cite{hgz97}] is to smooth the initial displacement
field on either the orbit-crossing or the Jeans scale, whichever
is larger. However, while the Jeans-smoothing is physically motivated
in the sense it is meant to model the effect of finite
gas pressure, the other smoothing is more of a corrective measure to
deal with the inaccuracy of the ZA after orbit-crossing. It is a legitimate
concern whether such smoothing erases structures that might contribute
significantly to $f({N_{\rm HI}})$. The test against a hydrodynamic
simulation shown in Fig. 1 lends support to the use of the smoothed ZA.
However, this might be a lucky coincidence. Further tests are needed.

Third, there is no guarantee that in current structure formation models,
a given peak in optical depth has exactly the Voigt-profile shape,
and so standard profile-fitting algorithms might result in one
single peak being fitted by several small
profiles, which means the association of
the one single peak with one absorption line is not exact. 
The average effect on $f(N_{\rm HI})$ is hard to predict analytically, and
has to be checked through simulations. Let us now turn to the latest
work that has bearing on the above issues.

In two separate pieces of work presented in this conference 
([\cite{d97}] \& [\cite{b97}]), it is shown using new hydrodynamic simulations
that the slope of $f({\rm N_{\rm HI}})$ is roughly the same for
several cosmological models (of the order of $4$ for each). 
This might indicate the third worry mentioned above is
justified: that the nature of the Voigt-profile fitting
procedure might conspire to result in the same $f({\rm N_{\rm HI}})$ for
different models, contrary to what would be expected using a peak-counting
method.

However, it could also be the case that the models simulated above
do not span a large enough range (or have small enough small-scale power) 
to allow one to see the effect on $f({\rm N_{\rm HI}})$.
A different line of attack is developed in [\cite{gh97}], in which
the Ly$\alpha$ forest is simulated using a PM code, with the Poisson
solver modified to compute an effective potential due to pressure
in addition to that due to gravity. A related approach is advocated by 
[\cite{pmk95}]. This method produces results in good agreement with 
hydrodynamic simulations for the low $N_{\rm HI}$ Ly$\alpha$ forest. It
avoids the problem of the uncertain smoothing scale for the ZA.
It also allows simulations of a large number of models
with relatively modest computer expense. 
This is undertaken by
[\cite{g97}], who finds appreciable differences in the slope of
$f({\rm N_{\rm HI}})$ among a set of $25$ cosmological models,
and concludes the normalization of the power spectrum at a particular
scale can indeed be constrained by $f({\rm N_{\rm HI}})$, confirming the prediction of 
[\cite{hgz97}]. It is also shown that the peak-counting method in real space
compares favorably with Voigt-profile-fitting as a way of finding
$f({\rm N_{\rm HI}})$. However, a few caveats have to be kept in mind. 
First, in [\cite{g97}], only a narrow range of $N_{\rm HI}$ 
($10^{13}-10^{14}\, {\rm cm^{-2}}$) is used to arrive at the above 
conclusions. Second, the correction introduced due to the finite
box size should be checked using larger simulations.
Third, the observed $f({\rm N_{\rm HI}})$ that the theoretical 
predictions are compared against is obtained using a different
profile-fitting algorithm from the one discussed in [\cite{g97}].
In general, the proper error-analysis is 
probably highly algorithm-dependent, when one attempts to use the 
observed $f({\rm N_{\rm HI}})$ to constrain the power spectrum.

The author thanks Nickolay Gnedin and Yu Zhang with whom
some of the work reviewed here has been done. Support by the DOE and
by the NASA (NAG5-2788) at Fermilab is gratefully acknowledged.


\begin{figure}
\centerline{\vbox{
\psfig{figure=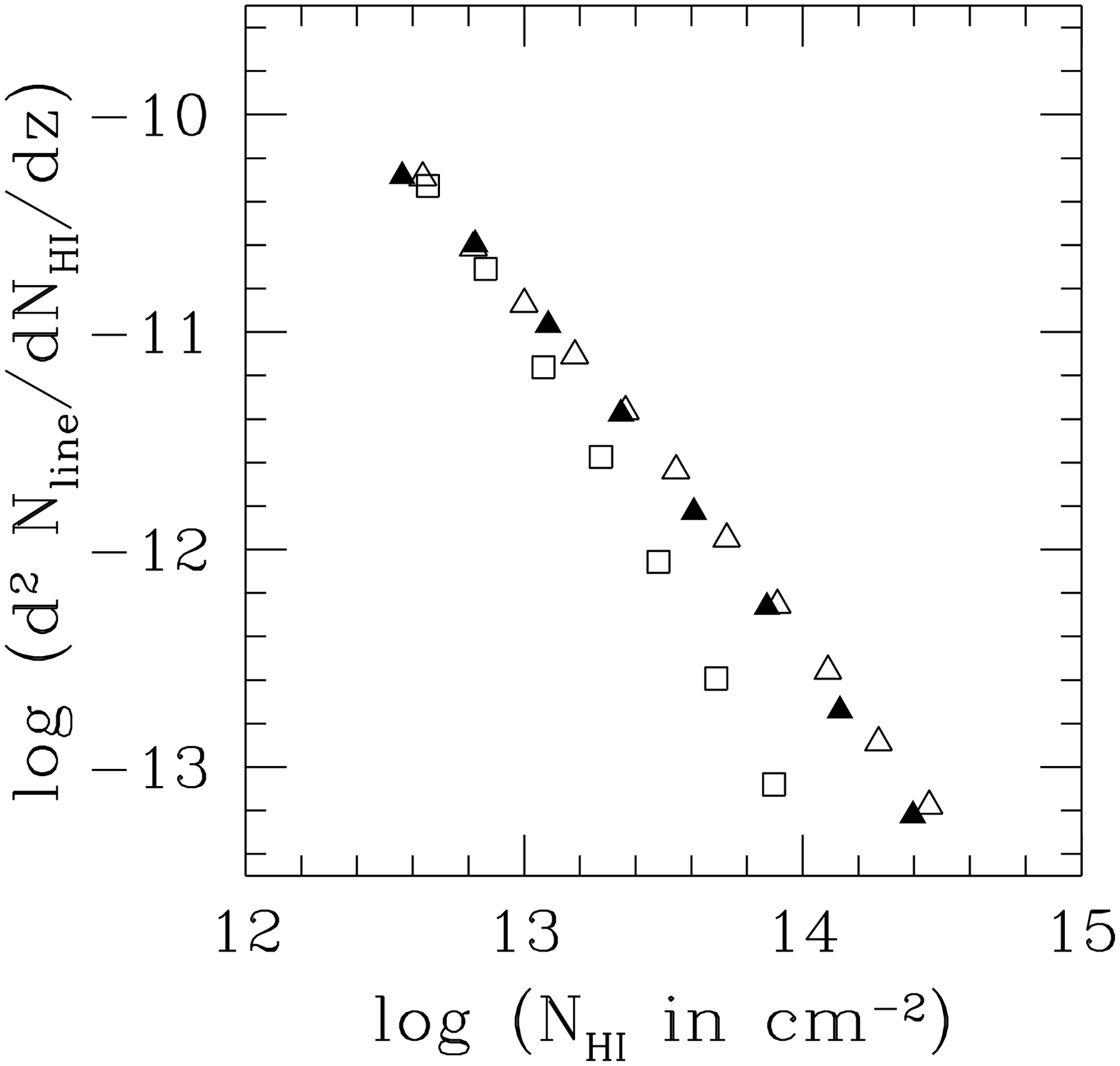,height=10.0cm,width=10.0cm}
}}
\caption[]{The column density distributions for a CDM model ({\it triangle})
and a CHDM model ({\it square}). Open/filled symbols are obtained using the ZA/hydrodynamic simulation respectively. See [\cite{hgz96}] \& [\cite{hgz97}] for details.
}
\end{figure}

\vfill
\end{document}